\documentclass[aps,prl,showpacs,twocolumn,floats,epsfig,pdflatex]{revtex4}
\usepackage{amssymb}
\usepackage{amsbsy}
\usepackage{amsmath}
\usepackage{epsfig}
\usepackage{graphicx}
\begin{document}

\title {Dynamic freezing of strongly correlated ultracold bosons}

\author{S. Mondal$^{(1)}$, D. Pekker$^{(2)}$, and K. Sengupta$^{(1)}$}

\affiliation{$^{(1)}$ Theoretical Physics Department, Indian
Association for the Cultivation of Science, Jadavpur,
Kolkata-700032, India. \\
$^{(2)}$ Department of Physics, California Institute of Technology,
Pasadena, California-91125, USA.}

\date{\today}

\begin{abstract}

We study the non-equilibrium dynamics of ultracold bosons in an
optical lattice with a time dependent hopping amplitude $J(t)=J_0
+\delta J \cos(\omega t)$ which takes the system from a superfluid
phase near the Mott-superfluid transition ($J= J_0+\delta J$) to a
Mott phase ($J=J_0-\delta J$) and back through a quantum critical
point ($J=J_c$) and demonstrate dynamic freezing of the boson
wavefunction at specific values of $\omega$. At these values, the
wavefunction overlap $F$ (defect density $P=1-F$) approaches unity
(zero). We provide a qualitative explanation of the freezing
phenomenon, show it's robustness against quantum fluctuations and
the presence of a trap, compute residual energy and superfluid order
parameter for such dynamics, and suggest experiments to test our
theory.

\end{abstract}


\maketitle

Theoretical study of non-equilibrium dynamics in closed quantum
systems has seen great progress in recent years \cite{rev1} mainly
due to the possibility of realization of such dynamics using
ultracold atom in optical lattices \cite{bloch1,exp1}. For bosonic
atoms, such systems are well described by the Bose-Hubbard model
with on-site interaction strength $U$ and nearest neighbor hopping
amplitude $J$ \cite{bosepapers1,bosepapers2}. Several theoretical
studies have been carried out on the quench and ramp dynamics of
this model \cite{koll1,anatoly1,gppapers1,ehud1,others1,sengupta1};
some of them have also received support from recent experiments
\cite{exp1}. In contrast, studies on periodically driven closed
quantum systems have been undertaken in the past mainly on driven
two-level systems \cite{rev2,per1} or on weakly interacting or
integrable many-body systems which can be modeled by them
\cite{per2,das1}. Among these, Ref.\ \cite{das1} has predicted
freezing of the time-averaged value of the order parameter
(magnetization) of an periodically driven one-dimensional (1D) Ising
or XY model, when the temporal average is performed over several
drive cycles, at specific drive frequencies. Such a freezing occurs
in the high frequency regime and exhibits non-monotonic dependence
on the drive frequency. However, to the best of our knowledge, the
phenomenon of dynamic freezing has never been demonstrated for
dynamics involving a single drive cycle and/or for non-integrable
quantum systems. Recent studies of periodic dynamics of the
Bose-Hubbard model have not addressed this issue \cite{gil1,susan1}.

In this work, we demonstrate, via designing a periodic driving
protocol, that the periodically driven Bose-Hubbard model may
exhibit dynamic freezing of the boson wavefunction
$|\psi(t=0)\rangle = |\psi(t=T)\rangle$ for specific values of the
drive frequencies $\omega=2\pi/T$. Our driving protocol constitutes
a time-dependent hopping amplitude of the bosons $J(t)= J_0 +\delta
J \cos(\omega t)$ with $J_0$ and $\delta J$ chosen such that the
drive takes the system from a superfluid (SF) ($J=J_0+\delta J$) to
the Mott insulator (MI) state ($J=J_0-\delta J$) and back through
the tip of the Mott lobe where $\mu=\mu_{\rm tip}$. We demonstrate,
using mean-field theory, that such a freezing phenomenon derives
from quantum interference of the dynamic phases acquired by the
bosons and compute the defect formation probability $P=1-F$ (where
$F = |\langle \psi(t=0)|\psi(t=T)\rangle|^2$ is the wavefunction
overlap), the superfluid order parameter $\Delta(T)= \langle
\psi(T)|b|\psi(T)\rangle$ (where $b$ denotes the boson annihiliation
operator), and the residual energy $Q= E(t=T)-E_G$ (where $E(t=T)$
is the energy of the system at the end of the drive cycle and $E_G$
is the initial ground state energy) as a function of $\omega$. We
also show, via inclusion of quantum fluctuation by a projection
operator approach \cite{sengupta1} and numerical mean-field study of
a trapped boson system that the freezing phenomenon is qualitatively
robust against quantum fluctuations and the presence of a trap. We
note that such a freezing behavior has two novel characteristics
which distinguishes it from its counterpart in Ref.\ \cite{das1}.
First it does not need high frequencies as $\hbar \omega/U \ll1$
throughout the range of $\omega$ where the freezing occurs. Second,
it occurs for a single cycle of the drive and does not need
averaging over several cycles. Such a dynamic freezing phenomenon
has not been studied in the context of closed quantum systems; our
work therefore constitutes a significant advance in our
understanding of periodic dynamics of closed non-integrable quantum
systems.

The Hamiltonian describing a system of ultracold bosonic atoms
confined by a trap and in an optical lattice is given by
\begin{eqnarray}
{\mathcal H} &=& \sum_{\langle {\bf r},{\bf r'}\rangle} -J  b_{{\bf
r}}^{\dagger} b_{{\bf r'}} + \sum_{{\bf r}} [-\mu_{\bf r} {\hat
n}_{{\bf r}} + \frac{U}{2} {\hat n}_{{\bf r}}({\hat n}_{{\bf r}}-1)]
\label{ham1}
\end{eqnarray}
where $\mu_{\bf r}$ denotes the chemical potential at site ${\bf
r}$, ${\bf r'}$ denotes one of the $z$ nearest neighboring sites of
${\bf r}$, and ${\hat n}_r= b_r^{\dagger} b_r$. In the absence of a
trap, $\mu_{\bf r}=\mu$ for all sites and for $zJ\ll U$, the ground
state of the model is a MI state with ${\bar n}$ bosons per site
with ${\bar n}=1$ for $0\le \mu/U \le 1$. For $zJ \gg U$, the bosons
are delocalized and the system, for $d\ge 2$, is in a SF state. In
between, at $J=J_c$, the system undergoes a SF-MI transition. The
equilibrium phase diagram of the model constitutes the well-known
Mott lobe structure \cite{bosepapers1,bosepapers2}.

To obtain an semi-analytic insight to the freezing phenomenon, we
first analyze the periodically driven Bose-Hubbard model in the
absence of a trap and within mean-field approximation. The
time-dependent mean-field Hamiltonian is given by
\begin{eqnarray}
{\mathcal H}_{{\rm mf}} &=& \sum_{{\bf r}} [-\mu {\hat n}_{{\bf r}}
+ \frac{U}{2} {\hat n}_{{\bf r}}({\hat n}_{{\bf r}}-1)] +\left
(\Delta'_r(t) b_r^{\dagger} + {\rm h.c} \right),\label{hammf}
\end{eqnarray}
where $\Delta'_r(t) = -J(t) \sum_{\langle r'\rangle} \langle b_{r'}
\rangle$ and $\Delta^{'}_0= \Delta'(t=0)$. Within homogeneous
mean-field theory, the Gutzwiller wavefunction for the bosons reads
$|\psi({\bf r},t)\rangle_{{\rm mf}}= \prod_{{\bf r}} \sum_n f_n(t)
|n\rangle$ \cite{kot1}. The Schrodinger equation $i
\partial_t |\psi({\bf r},t)\rangle_{{\rm mf}} = {\mathcal H}_{{\rm
mf}}(t) |\psi({\bf r},t)\rangle_{{\rm mf}}$ yields the
time-dependent mean-field equations for $f_{n}(t) \equiv f_n$:
\begin{figure} \includegraphics[width=\linewidth]{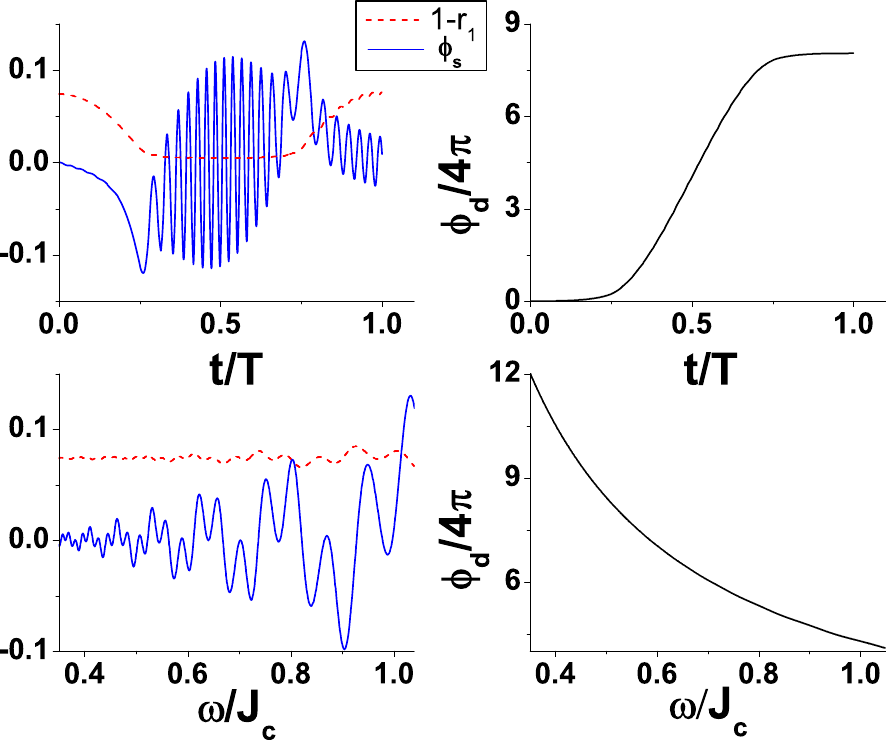}
\caption{(Color online) Top left panel: Plot of $1-r_1(t)$ (red
dashed line) and $\phi_s(t)$ (blue solid line) as a function of $t$.
Bottom left panel: Variation of $1-r_1(T)$ (red dashed line) and
$\phi_s(T)$ (blue solid line) with $\omega$. Top (Bottom) right
panel: Plot $\phi_d(t)$ ($\phi_d(T)$) as a function of $t$
($\omega$). For all plots $J_0=1.05 J_c$, $\delta J=0.35 J_c$, and
$\mu=0.414U$.} \label{fig1}
\end{figure}
\begin{eqnarray}
(i \partial_t -E_n)f_n  = {\tilde \Delta}(t) \sqrt{n} f_{n-1} +
{\tilde \Delta}^{\ast}(t) \sqrt{n+1} f_{n+1}, \label{mfteq1}
\end{eqnarray}
where ${\tilde \Delta}(t) = -z J(t) \sum_n \sqrt{n} f_{n-1}^{\ast}
f_n$, and $E_n= -\mu n +U(n-1)n/2$. In what follows, we shall choose
$J_0$ and $\delta J$ such that the ground state of $H_{\rm mf}$ with
$J=J_0 +\delta J$ is a SF state close to the QCP so that
$f_{n}(t=0)\simeq 0$ for $n\ge 3$ and $f_1(t=0) \gg f_0(t=0)\,,
f_2(t=0)$. As can be verified by explicit numerics, in this regime
$f_n(t)$ for $n\ge 3$ remains small for all $t$ during the dynamics
and can thus be neglected. The equations for $f_0$, $f_1$ and $f_2$
then reads (suppressing time dependence of $f_n(t)$ for clarity)
\begin{eqnarray}
i \partial_t f_0 &=& - zJ(t) [ |f_1|^2 f_0 + \sqrt{2}
f_2^{\ast} f_1^2] \label{mfteq2} \\
i \partial_t f_2 &=& E_2 f_2 - zJ(t) [2 |f_1|^2 f_2
+ \sqrt{2}  f_0^{\ast} f_1^2] \nonumber \\
i \partial_t f_1 &=& E_1 f_1 - zJ(t)  [(2 |f_2|^2 +|f_0|^2) f_1 + 2
\sqrt{2} f_1^{\ast} f_2 f_0 ]. \nonumber
\end{eqnarray}
From Eqs.\ \ref{mfteq2}, it is easy to see that $|f_n|$, for $n\le
2$, obeys the relation $\partial_t |f_0|^2 = \partial_t |f_2|^2 = -
\partial_t |f_1|^2/2$. Parameterizing $f_n=r_n(t) \exp[i \phi_n
(t)]$, with the choice that $\phi_n(0)=0$ and $r_n(0)=f_n(0)$, one
can write
%%%%%%%%%%%%%%%%%%%%%%%%%%%%%%%%%%%%%%%%%%%%%%%%%%%
\begin{figure} \includegraphics[width=\linewidth]{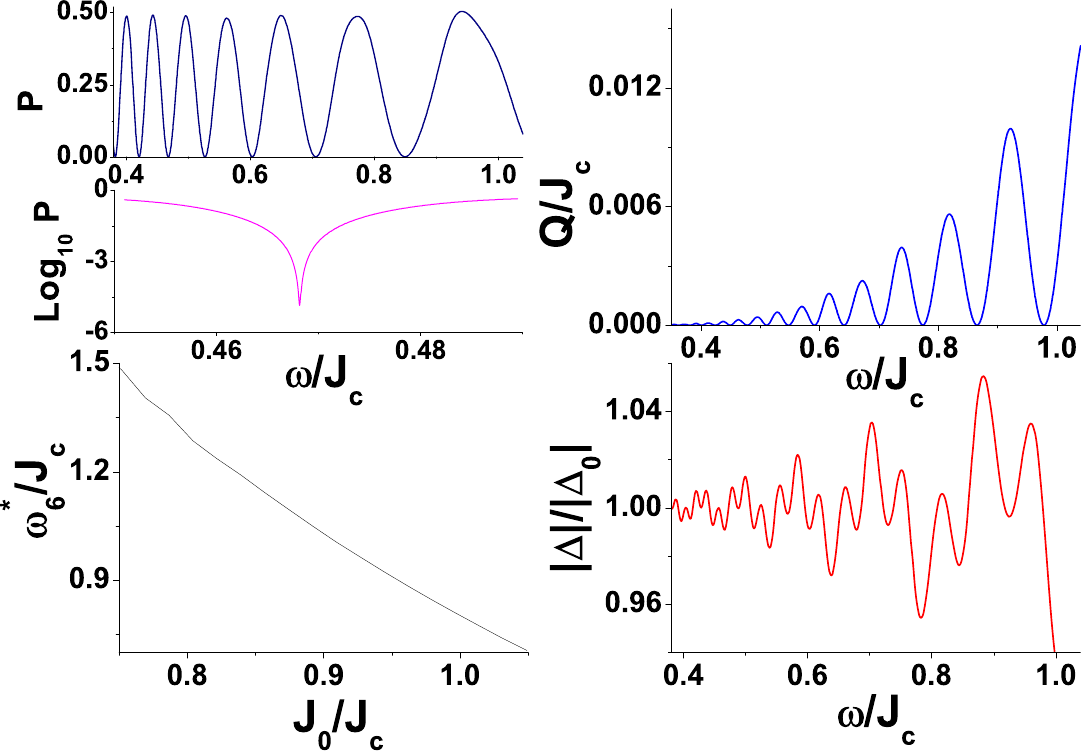}
\caption{(Color online) Top left panel: Plot of the defect density
$P$ as a function of $\omega/J_c$ displaying freezing at
$\omega=\omega_m^{\ast}$. Inset shows ${\rm Log}_{10} P$ near
$\omega_9^{\ast} \simeq 0.47 J_c$. Bottom left panel: Variation of
$\omega_m/J_c$ as a function of $J_0/J_c$ for $m=6$. Right panels:
Plot of $Q$ and $|\Delta(T)|$ as a function of $\omega/J_c$. All
parameters are same as Fig.\ \ref{fig1}.} \label{fig2}
\end{figure}
%%%%%%%%%%%%%%%%%%%%%%%%%%%%%%%%%%%%%%%%%%%%%%%%%%
\begin{eqnarray}
r_{2[0]}^2(t) &=& -(r_1^2(t) -1)/2 +[-] \eta, \label{mfteq3}
\end{eqnarray}
where $\eta$ is a time independent parameter whose value is fixed by
the initial values $r_n$ \cite{comment1}. Note that $\eta$
represents the magnitude of the particle-hole asymmetry since
$r_0=r_2$ for $\eta=0$. Substituting Eq.\ \ref{mfteq3} in Eq.\
\ref{mfteq2}, we get
\begin{eqnarray}
\partial_t r_1 &=& -\sqrt{2} z J(t) \sin(\phi_s) r_1 g_0(r_1), \nonumber\\
\partial_t \phi_s &=& -U +z J(t)\left[ g_1(r_1) - g_2(r_1) \cos(\phi_s) \right],
\label{mfteq4} \\
\partial_t \phi_d &=& -U+2\mu + z J(t) r_1^2 \left[ 1 - 4 \sqrt{2}
\eta \cos(\phi_s)/g_0(r_1) \right],  \nonumber
\end{eqnarray}
where we have suppressed the time dependence of $r_1$ and
$\phi_{s(d)}$ for clarity, $ \phi_s = \phi_0+\phi_2-2\phi_1$ and
$\phi_d=\phi_2-\phi_0$ are the sum and differences of the relative
phases of the Gutzwiller wavefunction, and the functions
$g_{i}(r_1)$ are given by
\begin{eqnarray}
g_0(r_1) &=& \sqrt{(1-r_1^2)^2-4\eta^2}, \quad g_1(r_1)= 6 r_1^2-3
-2 \eta, \nonumber\\
g_2(r_1) &=& 2 \sqrt{2} \left[ r_1^2(r_1^2-1)/g_0(r_1)+g_0(r_1)
\right]. \label{fns}
\end{eqnarray}
We note that the first two of the equations in Eq.\ \ref{mfteq4} are
coupled equations describing the evolution of $r_1$ and $\phi_s$,
while the third describes the evolution of $\phi_d$ in terms of
$r_1$ and $\phi_s$. Furthermore, using a scaled variable $t'=\omega
t/(2\pi)$, we find that the relation between $r_1$ and $\phi_s$ can
be written as
\begin{eqnarray}
dr_1/d\phi_s &=&  \frac{-\sqrt{2} \sin(\phi_s) r_1
g_0(r_1)}{[g_1(r_1)-g_2(r_1)\cos(\phi_s)]- U/zJ(t')}. \label{rel1}
\end{eqnarray}
This allows us to symbolically write $\phi_s = \xi(r_1,t')$, where
$\xi$ is an unknown function, and thus establish an $\omega$
independent relation between $r$ and $\phi_s$ for any fixed $t'$.

Eqs.\ \ref{mfteq3}, \ref{mfteq4} and \ref{fns} constitute the
central result of this work. They constitute a complete description
of the evolution of $f_0$, $f_1$, and $f_2$ in the presence of the
periodic drive and provide an understanding of the freezing
phenomenon as follows. First, we find that a numerical solution of
Eq.\ \ref{mfteq4}, together with Eq.\ \ref{mfteq3}, allows us to
obtain $r_n$, $\phi_s$ and $\phi_d$ as a function of time. A plot of
$1-r_1(t)$ and $\phi_s(t)$ as a function of $t$ for
$\omega/J_c=0.52$ is shown in the left panel of Fig.\ \ref{fig1}. We
find that $r_1$ changes appreciably when $J(t)$ is close to $J_c$;
however, $r_1(T) \simeq r_1(0)$ at the end of the evolution. Note
that this also implies, via Eq.\ \ref{mfteq3}, that $r_{2(0)}(0)
\simeq r_{2(0)}(T)$. The bottom left panel of Fig.\ \ref{fig1} shows
that the relation $r_1(T) \simeq r_1(0)$ holds for a significant
range $\omega/J_c \le 0.8$. Second, we note that, $\phi_s(t)$
undergoes rapid oscillation when $J(t) \le J_c$; however, it also
comes back close to it's initial value at the end of the drive:
$\phi_s(T) \simeq \phi_s(0)$. Since $\phi_s$ and $r_1$ satisfies a
$\omega$ independent relation, $\phi_s =\xi(r_1,t')$, we infer that
$\phi_s$ must remain close to its initial value for the same range
of $\omega$ for which $r_1(T) \simeq r_1(0)$; this is verified
numerically in the bottom left panel of Fig.\ \ref{fig1}. Finally,
we note, from the right panels of Fig.\ \ref{fig1}, that $\phi_d(T)$
is a monotonic function of $\omega$. Thus we may define
$\omega=\omega_m^{\ast}$ for which $\phi_d(T) \simeq 4 \pi m$ ($m$
being an integer). Together with the fact that $r_1(T) \simeq
r_1(0)$ and $\phi_s(T) \simeq \phi_s(0)=0$, we find that at $\omega
=\omega_m^{\ast}$, both the relative phases satisfy
$\phi_2-\phi_1=-(\phi_0-\phi_1) \simeq 2\pi m$ leading to
$|\psi_{\rm mf}(T)\rangle \simeq |\psi_{\rm mf}(0)\rangle$ up to a
global phase. This constitutes the dynamics freezing of
$|\psi\rangle_{\rm mf}$.

\begin{figure} \includegraphics[width=\linewidth]{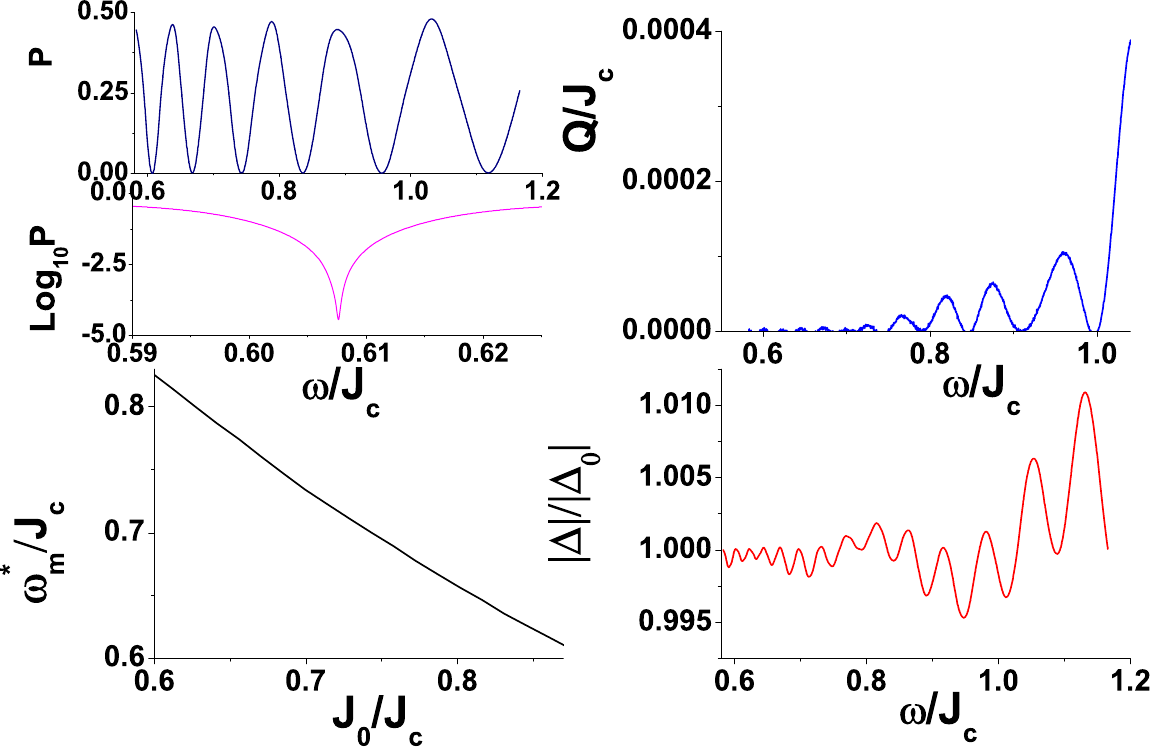}
\caption{(Color online) Similar plots as in Fig.\ \ref{fig2} but
with $J_0[\delta J]=0.87 [0.58]J_c$ and computed using the
projection operator approach displaying robustness of the freezing
phenomenon against quantum fluctuations. See text for details.}
\label{fig3}
\end{figure}

To obtain an accurate estimate of the degree of freezing, we compute
the defect density $P=1-F=1-|\langle \psi_{\rm mf}(T)|\psi_{\rm
mf}(0)\rangle|^2$. The plot of $P$ as a function of $\omega$ clearly
shows that $P \to 0$ at $\omega=\omega_m^{\ast}$. A plot of ${\rm
Log}_{10} P$ vs $\omega$ near $\omega_9^{\ast}/J_c \simeq 0.47$,
shown in the top left panel of Fig. \ref{fig2} reveals that $P \sim 10^{-6}$ 
indicating that the overlap, up to a global
phase, is exact within our numerical accuracy. We have checked for
all $\omega_m^{\ast} \le 0.8 J_c$, $P < 10^{-4}$ which
indicates a near perfect freezing. We also compute the residual
energy $Q(T)$ and the SF order parameter
\begin{eqnarray}
\Delta &=& r_1 e^{i\phi_d/2} \left(r_0 e^{-i \phi_s/2} + \sqrt{2}
r_2 e^{i\phi_s/2}\right), \label{op1}
\end{eqnarray}
at $t=T$ as a function of $\omega$. We find from Eq.\ \ref{op1} that
$|\Delta|$ is independent of $\phi_d$. Thus $|\Delta(T)|/|\Delta_0|
$ and $Q/U$ (which can also be shown to be independent of $\phi_d$)
remain close to unity and zero respectively over the entire range of
$\omega/J$ for which $r_1$ and $\phi_s$ remain close to their
initial values as shown in right panels of Fig.\ \ref{fig2}. Such a
behavior distinguishes these quantities from $P$ which depends on
$\phi_d$ and hence vanishes at discrete $\omega_m$. Finally, we find
that for all values of $J_0$ shown in bottom left panel of Fig.\
\ref{fig2}, there is an appreciable range of $\omega/J_c$ within
which the freezing phenomenon occurs and that $\omega_m^{\ast}$
decreases monotonically as a function of $J_0$ over this range.

Next, we study the effect of quantum fluctuations on the freezing
phenomenon. We incorporate such fluctuations by using a projection
operator method developed in Ref.\ \onlinecite{sengupta1} which
provides an accurate treatment of dynamics with fluctuations for
$zJ(t)/U \ll 1$. The idea behind this approach, as detailed in Ref.\
\onlinecite{sengupta1}, is to introduce a projection operator
$P_{\ell}= |n_0\rangle \langle n_0|_{\bf r} \times |n_0\rangle
\langle n_0|_{\bf r'}$ which lives on the link $\ell$ between the
neighboring sites ${\bf r}$ and ${\bf r'}$ of the lattice. Using
$P_{\ell}$, one can write the boson hopping term as $T' =
\sum_{\langle {\bf r} {\bf r'} \rangle} -J(t) b_{{\bf r}}^{\dagger}
b_{\bf r'} = \sum_{\ell} T'_{\ell} = \sum_{\ell} [ (P_{\ell}
T'_{\ell} + T'_{\ell} P_{\ell}) + P_{\ell}^{\perp} T'_{\ell}
P_{\ell}^{\perp}]$ where $P_{\ell}^{\perp}=(1-P_{\ell})$. In the
strong-coupling regime where $zJ(t)/U \ll 1$, the term
$T_{\ell}^{'0}[J] = (P_{\ell} T'_{\ell} +T'_{\ell} P_{\ell})$, at
any instant, represents hopping processes which takes the system out
of the instantaneous low-energy manifold. Thus one can devise a
time-dependent canonical transformation via an operator $S \equiv
S[J(t)]= \sum_{\ell} -i[P_{\ell},T'_{\ell}]/U$ which eliminates
$T_{\ell}^{'0}[J(t)]$ up to first order in $J(t)/U$ and leads to the
effective instantaneous time-dependent Hamiltonian
$H^{\ast}=\exp(-iS[J(t)]) {\mathcal H} \exp(iS[J(t)])$. Such a
canonical  transformation is equivalent to a transformation on the
system wavefunction $|\psi\rangle$: $|\psi'\rangle = \exp(-i
S[J(t)]) |\psi\rangle$. We note that $|\psi\rangle$ and
$|\psi'\rangle$ coincides for $J=0$ which leads us to the natural
choice $|\psi'\rangle = \prod_{\bf r} \sum_n f_{\bf r}^n(t) |n_{\bf
r}\rangle$. Note that $|\psi\rangle$ is not of Gutzwiller form; it
involves spatial correlation due to $\exp(iS[J(t)])$ factor. The
instantaneous energy of the system is given by $E[\{f_{\bf
r}^n(t)\}] = \langle \psi|{\mathcal H} | \psi \rangle = \langle
\psi'|H^{\ast}| \psi'\rangle + {\rm O}(J(t)^3/U^3)$ and includes
$O(J^2/U^2)$ quantum fluctuation corrections. As shown in Ref.\
\cite{sengupta1}, this formalism allows one to describe the dynamics
of the bosons by solving for the Schrodinger equation for
$|\psi'\rangle$:
\begin{eqnarray}
(i \hbar \partial_t + \partial S[J(t)]/\partial t) |\psi'\rangle =
H^{\ast}[J(t)] |\psi' \rangle. \label{sch1}
\end{eqnarray}
Using the expression of $|\psi'\rangle$ and $E[\{f_n\}]$, one can
convert Eq.\ \ref{sch1} to a set of equations for $\{f_{{\bf
r}}^n(t)\}$ \cite{sengupta1}. Defining $\varphi_n = \sqrt{n+1}
f_n^{\ast} f_{n+1}$, one gets
\begin{eqnarray}
i \hbar \partial_t f_n  &=& \delta E[\{f_n(t)\};J(t)]/ \delta
f_n^{\ast} + \frac{i z \hbar}{U} \frac{\partial J(t)}{\partial t}
\nonumber\\
&& \times \Big( \sqrt{n} f_{n-1}
\Big[\delta_{n\bar{n}}\varphi_{\bar{n}}
- \delta_{n,\bar{n}+1}\varphi_{\bar{n}-1}\Big] \nonumber \\
&& + \sqrt{n+1} f_{n+1} \Big[ \delta_{n\bar{n}}\varphi_{\bar{n}-1}^*
- \delta_{n,\bar{n}-1} \varphi_{\bar{n}}^* \Big] \Big).
\label{proj2}
\end{eqnarray}

\begin{figure} \includegraphics[width=\linewidth]{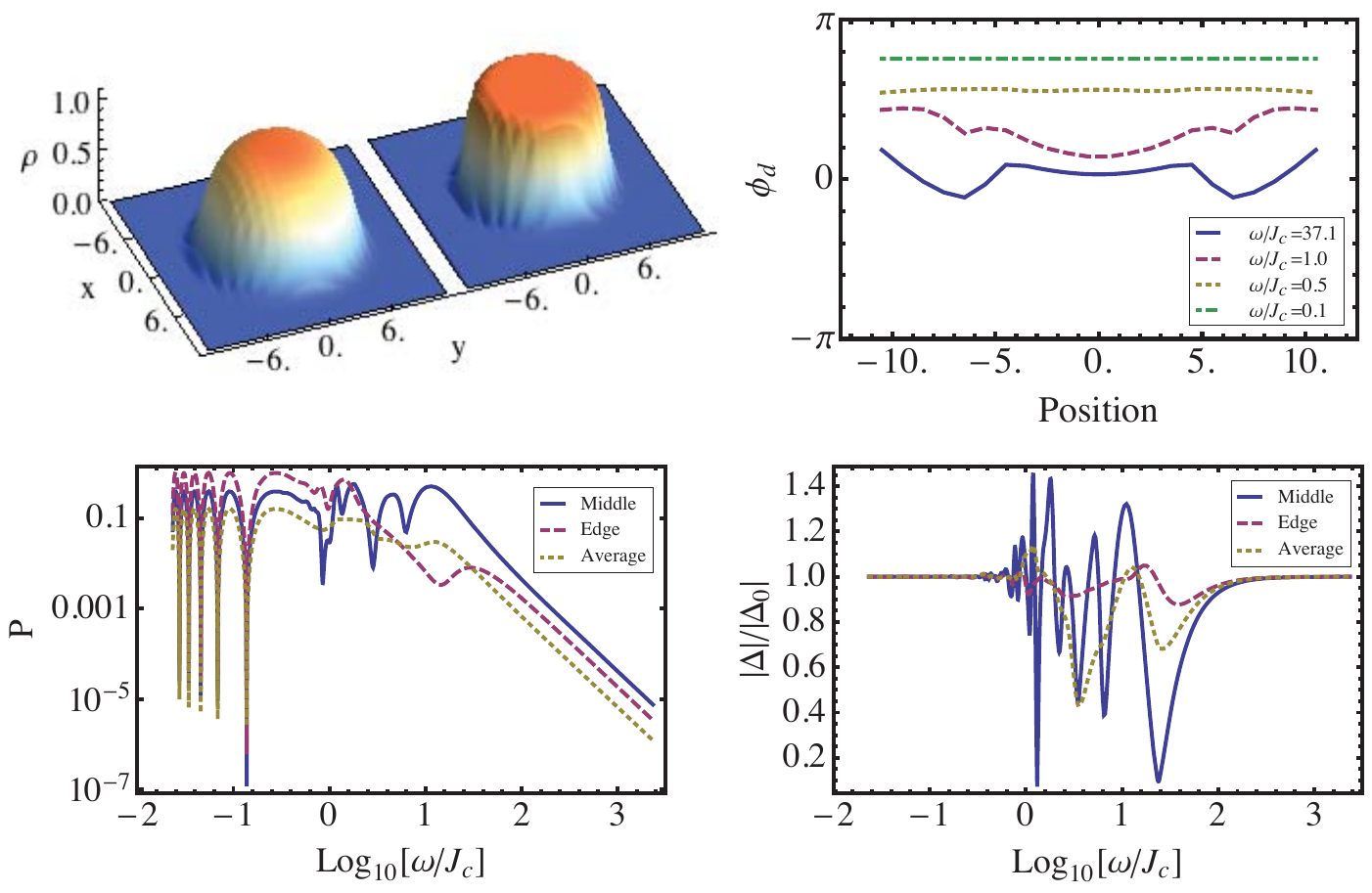}
\caption{(Color online) Top left panel: Plot of the boson density
profile in the trap at $t=0$ and $t=T$. Top right panel: Variation
of relative phase $\phi_d(T)$ as a function of the position (along
$y=0$) in the trap displaying coherent evolution of the bosons.
Bottom panels: Plot $P$ and $|\Delta(T)|/|\Delta_0| $ as a function
$\omega$ displaying freezing phenomenon at $\omega_m^{\ast}$. For
all plots $J_0=0.04 U$, $\delta J=0.015 U$, $J_c=0.041 U$, and
$\mu_0 =0.415 U$.} \label{fig4}
\end{figure}

A numerical solution of Eq.\ \ref{proj2} yields $f_n(t)$ and hence
$|\psi'\rangle$ using which one can compute $ |\psi(t)\rangle =
\exp[iS] |\psi'(t)\rangle$ perturbatively to ${\rm O}[J(t)^2/U^2]$.
Similarly, expectation value of any operator $O$ at any instant $t$
can be calculated in terms of $|\psi'(t)\rangle$: $\langle O\rangle
= \langle \psi'(t)|e^{-iS} O e^{iS}| \psi'(t)\rangle = \langle
\psi'(t)|O| \psi'(t)\rangle - \langle \psi'(t)|[iS,O]|
\psi'(t)\rangle + ...$, where the ellipsis indicate higher order
terms in $J(t)/U$. Note that the second term in the expression
originates from quantum fluctuation and modifies mean-field result
(first term). Using the above-mentioned procedure detailed in Ref.\
\cite{sengupta1}, we compute $P(T)$, $Q(T)$ and $|\Delta(T)|$ as
shown in Fig.\ \ref{fig3}. We find that key effects of the quantum
fluctuations is to change numerical values of $\omega_m^{\ast}$ and
the precise range of $\omega$ over which freezing occurs; however
the mean-field results hold qualitatively in the sense that $P \to
0$ for several $\omega_m^{\ast}$ with ${\rm Log}_{10} P \le -4$ for
all $\omega_m^{\ast}$. Further $\omega_m^{\ast}$ also decreases
monotonically with $J_0$ as shown in left bottom panel of Fig.\
\ref{fig3} for $\omega^{\ast} \simeq 0.6 J_c$.

Finally, we consider the effect of a harmonic trap on the freezing
phenomenon. For this part, we numerically solve Eq.\ \ref{mfteq1}
for $d=2$ with $\mu_{\bf r}= \mu_0 +0.01 U[({\bf r}_x-1/2)^2+({\bf
r_y}-1/2)^2]$, for $N_0=576$ sites (linear dimension $24$) and with
fixed total particle number $N_0$. We choose the trap parameters so
that the ground state of the bosons in center of the trap at $t=T/2$
is MI phase with ${\bar n}=1$. The evolution of the density profile
of the bosons is shown in the top left panel of Fig.\ \ref{fig4} for
$t=0$ (left) and $t=T/2$ (right). The top right panel indicates
evolution of $\phi_d$ as a function of the position of the bosons in
the trap along the line $y=0$. The plot indicates that for all
$\omega \le J_c$, $\phi_d$ evolves coherently with negligible
spatial variation. The plots for $\phi_s$ and $r_1$ are similar in
nature; thus, we expect the boson evolution to have the same
qualitative properties as that found within a homogeneous mean-field
approach. A plot of $P$ ($|\Delta(T)|/|\Delta_0| $) as a function of
${\rm Log}_{10}(\omega/J_c)$ in the lower left (right) panels of
Fig.\ \ref{fig4} confirms this expectation. We find that the main
effect of the trap is to push the freezing phenomenon to lower
frequencies leaving its qualitative nature unchanged. The largest
freezing frequency occurs at $\simeq 0.2 J_c$ which is large
compared to frequencies $\simeq 0.05 J_c$ where momentum conserving
boson pair production at finite momenta, which is not captured
within mean-field theory, is expected to become significant
\cite{dp1}. Fig.\ \ref{fig4} also demonstrates that the freezing
phenomenon disappears at higher drive frequencies where the trapped
bosons do not evolve coherently leading to spatial variation of
$\phi_d$.

For experimental verifications of our work, we suggest interference
of two bosonic condensates in the presence of an optical lattice,
near the QCP which are separated after creation by a double-well
potential and allowed to evolve separately for a fixed holdout time.
It is well known that recombination of such separated condensates
can act as a readout scheme for their relative phases \cite{exp2}.
We propose such a readout when one of the condensates is driven
periodically with a frequency $\omega$ during the holdout for a
single period $T=2\pi/\omega$. Our specific prediction is that the
relative phase measured for such a drive with
$\omega=\omega_m^{\ast}$ is going to match the phase without any
drive indicating dynamic freezing. For all such experiments one
needs to estimate a optimal temperature $T_0$ at which they can be
carried out. The typical value of $U$ deep inside the Mott phase is
$\simeq 2 {\rm KHz} = 200$nK leading to a melting temperature of
$T_m \simeq 0.2 U = 40$nK for $d=3$. The SF phase near the Mott tip
has a coherence temperature of $T_c \simeq z J_c \simeq 35$nK
\cite{gerbier1}. Thus a temperature of a few nano-Kelvins ($T_0 \ll
T_m, T_c$), which is currently within the experimental reach, would
be ideal for testing our prediction.

In conclusion, we have demonstrated that periodic dynamics of the
ultracold bosons described by the Bose-Hubbard model leads to
dynamic freezing of the Boson wavefunction at specific drive
frequencies which are determined by the condition $\phi_d(T)= 4 \pi
m$. The freezing phenomenon is qualitatively robust against the
presence of the trap and quantum fluctuations; it manifests itself
at discrete drive frequencies $\omega_m^{\ast} \le J_c$ via presence
of dips in the defect density and can be detected by suitable
interference experiments.

SM and KS thanks K. Ray for several stimulating discussions. KS
thanks DST for support through grant SR/S2/CMP-001/2009. DP
acknowledges support from the Lee A. DuBridge fellowship.


\begin{thebibliography}{99}


\bibitem{rev1}A. Polkovnikov {\it et al.}, \rmp {\bf 83}, 863 (2011); J.
Dziarmaga, Adv. Phys. {\bf 59}, 1063 (2010).

\bibitem{bloch1} M. Greiner, {\it et al.}, Nature {\bf 415}, 39 (2002);
C. Orzel {\it et al.}, Science {\bf 291}, 2386 (2001); Kinoshita,
T., T. Wenger, and D. S. Weiss, Nature {\bf 440}, 900 (2006); L. E.
Saddler {\it et al.}, Nature {\bf 443}, 312 (2006).

\bibitem{exp1} W.S. Bakr {\it et al.}, Science {\bf 329}, 547 (2010).

\bibitem{bosepapers1} D. Jaksch {\it et al.}, Phys. Rev. Lett.
{\bf 81}, 3108 (1998); K. Sengupta and N. Dupuis, \pra {\bf 71},
033629 (2005); J. Freericks {\it et al.}, \pra {\bf 79}, 053631
(2009).

\bibitem{bosepapers2} W. Krauth and N. Trivedi, Europhys. Lett.
{\bf 14}, 627 (1991); B. Caprogrosso-Sansone, N. Prokofiev, and B.
V. Svistunov, Phys. Rev. B {\bf 75}, 134302 (2007).

\bibitem{koll1} C. Kollath, A. Lauchli, and E. Altman, \prl {\bf 98},
180601 (2007).

\bibitem {anatoly1} C. De Grandi, V. Gritsev, A. Polkovnikov,
Phys. Rev. B {\bf 81}, 224301 (2010); C. De Grandi, R. A. Barankov,
and A. Polkovnikov, Phys. Rev. Lett. {\bf 101}, 230402 (2008); C. De
Grandi, V. Gritsev, and A. Polkovnikov, Phys. Rev. B {\bf 81},
012303 (2010); C. de Grandi and A. Polkovnikov, {\it Quantum
Quenching, Annealing and Computation}, Eds. A. Das, A. Chandra and
B. K. Chakrabarti, Lect. Notes in Phys., {\bf 802} (Springer,
Heidelberg 2010).

\bibitem{gppapers1} A. Polkovnikov, \pra {\bf 66}, 053607 (2002);
A. Polkovnikov and V. Gritsev, Nat. Phys. {\bf 4}, 477 (2006).

\bibitem{ehud1}E. Altman and A. Auerbach, Phys. Rev. Lett. {\bf 89},
250404 (2002)

\bibitem{others1}R. Schutzhold {\it et al.}, \prl {\bf 97}, 200601
(2006); J. Wernsdorfer {\it et al.} \pra {\bf 81}, 043620 (2010).

\bibitem{sengupta1} C. Trefzger and K. Sengupta, \prl {\bf 106}
095706 (2011); A.Dutta, C. Trefzger, and K. Sengupta,
arXiv:1111.5085 (unpublished).

\bibitem{rev2} S.N. Shevchenko, S. Ashhab, and F. Nori,
Phys. Rept. {\bf 492}, 1 (2010).

\bibitem{per1} S.N. Shevchenko, S. Ashhab, and F. Nori, \prb {\bf
85} 094502 (2012).

\bibitem{per2} L-K Kim, J-N Fuchas and G. Montambaux,
arXiv:1201.1479 (unpublished); R. de Gail {\it et al.}
arXiv:1203.1262 (unpublished).

\bibitem{das1} A. Das \prb {\bf 82}, 172402 (2010); S. Bhattacharya,
A. Das, and S. Dasgupta, arXiv:1112.6171 (unpublished).

\bibitem{gil1} A. Robertson, V.M. Galitski, and G. Refael, \prl {\bf
106}, 165701 (2011)

\bibitem{susan1} S. Pielawa, Phys. Rev. A {\bf 83}, 013628 (2011).

\bibitem{kot1} D. Rokhsar and B. G. Kotliar, \prb{\bf 44}, 10328
(1991).

\bibitem{comment1} We note that within the single site homogeneous
mean-field theory, the system does not  exhibit freezing for
$\eta=0$. This behavior originates from the constraint of
conservation of particle number at each site and is not seen in
realistic systems with traps where only the total particle number is
conserved.

\bibitem{dp1} D. Pekker, B. Wunsch, T. Kitagawa, E. Manousakis, A. S. Sorensen, E. Demler {\it et al} (unpublished).

\bibitem{exp2} M. R. Andrews {\it et al.}, Science {\bf 275}, 637
(1997); T. Schumm {\it et al.}, Nat. Phys. {\bf 1}, 57 (2005); G.-B.
Jo {\it et al.}, \prl {\bf 98}, 180401 (2007).

\bibitem{gerbier1} F. Gerbier, Phys. Rev. Lett. {\bf 99}, 120405 (2007);
D.M. Weld {\it et al.}, Phys. Rev. Lett. {\bf 103}, 245301 (2009).

\end{thebibliography}
\end{document}